\documentclass[a4paper]{article}
\usepackage{amsmath,amssymb,amsfonts,epsfig}

\def\b#1{\mathbf{#1}}

\def\DS#1{$\displaystyle{#1}$}

\def\h#1{\widehat{#1}}

\def\lr{\leftrightarrow}

\def\o#1{\overline{#1}}

\def\op#1{\overset{\leftrightarrow}{#1}}
\def\p{\partial}
\def\px#1#2{\frac{\partial#1}{\partial#2}}
\def\pxx#1#2#3{\frac{\partial^2#1}{\partial#2\partial#3}}

\def\v#1{\overrightarrow{#1}}
\def\w#1{\widetilde{#1}}
\def\tr{\mathsf{Tr}}

\def\C{\mathcal{C}}

\def\F{\mathcal{F}}

\def\H{\mathcal{H}}

\def\O{\mathcal{O}}
\def\P{\mathcal{P}}
\def\Q{\mathcal{Q}}

\def\V{\mathcal{V}}

\def\NB{\mathbb{N}}

\begin{document}

\begin{titlepage}

\begin{flushright}
CPHT-RR008.0110\\
LPT-10-08\\
IPHT-T10/005
\end{flushright}

\vspace*{0.2cm}
\begin{center}
{\Large {\bf Virtual Compton Scattering  \\
off a Spinless Target in AdS/QCD 
}}\\[2 cm]

 {\bf Cyrille Marquet}~$^a$,  {\bf Claude Roiesnel}~$^b$ 
 {\bf and Samuel Wallon}~$^{ c,d}$\\[1cm]

 $^a$ {\it Institut de Physique Th\'eorique, CEA/Saclay, 91191
   Gif-sur-Yvette,
   France}\\[0.5cm]
 $^b$ {\it Centre de Physique Th\'eorique, \'Ecole Polytechnique,
   CNRS,
   91128 Palaiseau, France}\\[0.5cm]
 $^c$ {\it LPT, Universit\'e d'Orsay, CNRS, 91405 Orsay, France}\\ 
$^d$ {\it UPMC Univ. Paris 06, facult\'e de physique, 4 place Jussieu,
   75252 Paris Cedex 05, France}

\end{center}

\vspace*{2.0 cm}

\begin{abstract}
  We study the doubly virtual Compton scattering off a spinless target
  $\gamma^*P\to\gamma^*P'$ within the Anti-de Sitter(AdS)/QCD
  formalism.  We find that the general structure allowed by the
  Lorentz invariance and gauge invariance of the Compton amplitude is
  not easily reproduced with the standard recipes of the AdS/QCD
  correspondence.  In the soft-photon regime, where the semi-classical
  approximation is supposed to apply best, we show that the
  measurements of the electric and magnetic polarizabilities of a
  target like the charged pion in real Compton scattering, can already
  serve as stringent tests.
\end{abstract}

\end{titlepage}

\section{Introduction}
\label{INTRO}

The asymptotic conformal invariance of QCD has fostered a number of
theoretical attempts to push the original AdS/CFT duality \cite{MA98}
beyond its conjectured domain of validity.  The AdS/CFT correspondence
postulates more generally a relation, through a well-defined set of
prescriptions \cite{WI98, GKP98}, between weakly coupled string
theories living in the bulk of an anti-de Sitter (AdS) space and
strongly coupled conformally invariant field theories defined on its
boundary. The efforts within the so-called AdS/QCD approach rely on
the assumption that the AdS/CFT dictionary can still describe the
strong coupling regime of a confining gauge theory like QCD, despite
the breaking of conformal invariance, by computing correlation
functions within a semi-classical field theory formulated in a
five-dimensional AdS spacetime.

The various AdS/QCD models usually fall into two main categories. In
the so-called ``top-down'', or ``gauge/string'', approach one sticks
as much as possible to the original formulation of the AdS/CFT
correspondence, and starts from a superstring theory living on
$AdS_5\times M_5$ (where $M_5$ is some five-dimensional compact
manifold) and tries to derive an effective theory which describes the
low energy phenomena in QCD. The second, ``bottom-up'', or
``gauge/gravity'', approach puts aside the string-theoretic motivation
of the correspondence and starts from a phenomenological Lagrangian in
an appropriate five dimensional metric background which
incorporates as much as possible the known properties of QCD. Contrary
to some claims, the former approach is no more rigourous, in its
present stage, than the latter.

The simplest way to break conformal invariance is by introducing a
hard cut-off in AdS space which can be interpreted as an infrared mass
scale of the gauge theory. The first ``hard-wall'' model \cite{PS01}
was able to generate a power-law scaling of glueball elastic
scattering amplitudes at fixed angles. It is also possible to break
conformal invariance softly through a background dilaton field which
can be chosen \cite{KKSS} so as to reproduce the linear Regge behavior
of the meson trajectories.  Inside the bottom-up approach, it has
proved possible to reproduce qualitatively within both hard-wall and
soft-wall models the spectra of low-lying hadron states as well as
various decay constants and coupling constants
\cite{BOS02,TER05,HAM06,COL08}.

A frequent criticism claims that AdS/QCD models are nothing but some
kind of ``bag models'' since quite a few other phenomenological models
of QCD are able to reproduce the static parameters of hadronic states
at the same level of accuracy. In fact such criticisms are pretty
superficial because they ignore that AdS/QCD can incorporate in a
coherent framework several of these models (vector meson dominance,
1/N expansion, sum rules, $\cdots$, see \cite{EH09} for an up-to-date
review of the pros and cons of the AdS/QCD approach).

Moreover the AdS/QCD models can also be used to study in a completely
relativistic Lorentz invariant manner dynamical non-perturbative
aspects like hadronic form factors or structure functions. In
particular one can expect that conformal symmetry is most relevant for
describing the electromagnetic interactions of hadrons, since the
photon is a massless particle \footnote{e.g. there is a dynamical
  $O(4,2)$ symmetry which is sufficient to determine completely the
  spectrum and eigenfunctions of the relativistic hydrogen atom.}. A
very important process that reveals the internal structure of hadrons
is the deep inelastic scattering (DIS) of a highly energetic lepton
off a hadron.  This process was first investigated in a hard-wall
model in \cite{PS03}. Since then, there
have been a number of related studies within the different flavors of
AdS/QCD models \cite{HAT07,BAY08-1,BAY08-2,COR08,ALB08,HAT08,BAY08-3,
  GAO09,HAT09,LEV09,COR10,KOV09}.  However the semi-classical density
of states of the hard-wall or soft-wall models with canonical
dimensions does not agree \cite{PRSW} with the power-law behavior of
the structure functions observed at high energy.  It is presently far
from clear how a correct partonic description of hadrons can emerge
from the stringy corrections expected in the kinematical regime of
DIS.

On the other hand the AdS/QCD models provide a low-energy description
of the electromagnetic form factors which does agree with the
dimensional counting rules of hadrons up to a few GeV
\cite{GR07-1,GR07-2,BRO07,KL07,AC08,WARD08}.  Getting additional
information about the electromagnetic structure of hadrons requires
the study of four-point functions. There is one distinguished
electromagnetic process which supplies all experimentally accessible
information, namely Compton scattering.  There has been a lot of
theoretical work about Compton scattering.  Strong interactions can
significantly modify the amplitude but there is a low-energy theorem
\cite{LOW,GG54} which guarantees that the Born contribution dominates
near threshold.  The two leading orders of an expansion of the real
Compton scattering amplitude off a nucleon in terms of the frequency
of the photon are entirely given in terms of the charge, mass and
anomalous magnetic moment of the nucleon. For spinless targets like
the pion, there is no linear term in the energy of photons. This
theorem is based only on Lorentz invariance, gauge invariance and
crossing symmetry. Quadratic corrections to the Born terms (electric
and magnetic polarizabilities as well as generalized polarizabilities
with virtual photons) are not specified by symmetry arguments alone
and characterize non point-like elementary particles.  Measuring these
quantities allows to test the different models of strong interactions
(see \cite{PP95} for a review of the theoretical predictions on
charged and neutral pion polarizabilities).

The purpose of this work is to study the Compton amplitude off a
spinless target in the AdS/QCD formalism. We shall focus on the
kinematical region where the photons are soft. The semi-classical
approximation in the AdS/QCD duality should apply best in this region
and there exist experimental measurements of the pion polarizabilities
to compare with \cite{MAMI,COMPASS}. But we shall also study the
kinematical region with one deeply virtual photon (DVCS) since it is
straightforward to extract the corresponding structure functions in
the Bjorken limit.

The plan of the paper is as follows. We shall begin by a brief summary
of the standard lore about the general structure of the virtual
Compton amplitude off an unpolarized target. Next we shall describe
the generic soft-wall model we work with. In section \ref{4PT} we
shall explain how to calculate four-point functions in AdS/QCD and in
section \ref{SCALAR} we shall compute the Compton amplitude generated
by the minimal coupling of a bulk scalar field with a bulk U(1) gauge
field. We also compare our results with the recent hard-wall
calculation of \cite{GX09}. In the subsequent section we shall make
explicit the structure of this Compton amplitude in the deep inelastic
region, and in section \ref{DVCS} we shall clarify its Lorentz and
gauge-invariant structure in the DVCS kinematical region. Then we
shall extract the corresponding structure functions in the Bjorken
limit.  Section \ref{POL} is devoted to the calculation of the
polarizabilities of a spinless target. We shall comment on the
implications of our results in the conclusion.

\section{Virtual Compton amplitude off an unpolarized target}
\label{COMPTON}

The amplitude of the virtual Compton scattering, $\gamma^{\star}(q_1)
+ A(p_1) \rightarrow \gamma^{\star}(q_2) + A(p_2)$, where $A$ is a
spinless, or spin-averaged, hadron is defined through the off-forward
matrix element of the time-ordered product of two electromagnetic
currents,
\begin{gather}
  T_{\mu\nu} = i\int d^4x\,e^{iq\cdot x}\,\langle p_2
  \vert T\{J_{\mu}(x/2)J_{\nu}(-x/2)\}\vert p_1\rangle\,,\quad
  q=\frac{1}{2}(q_1+q_2)\,.
\end{gather}
In general, a two-to-two scattering amplitude depends on six
independent kinematical invariants, namely the external virtualities
$q_1^2, q_2^2, p_1^2, p_2^2$ and the usual Mandelstam variables
\begin{gather}
  s = (p_1+q_1)^2\,,\ t=(p_1-p_2)^2\,,\ u= (p_2-q_1)^2\,,
\end{gather}
obeying the constraint
\begin{gather}
  s + t + u = q_1^2 + q_2^2 + p_1^2 + p_2^2\,.
\end{gather}
It is convenient, for calculating the Compton form factors, to choose
$q_1$, $q_2$ and $p=p_1+p_2$ as the three independent momenta of the
process. At most thirteen independent tensors can contribute to the
Compton amplitude,
\begin{gather}
  \begin{split}
 g^{\mu\nu},\,p^{\mu}p^{\nu},\,q_1^{\mu}q_1^{\nu},\,
  q_2^{\mu}q_2^{\nu},\,q_1^{\mu}q_2^{\nu},\,q_2^{\mu}q_1^{\nu},\,
  p^{\mu}q_1^{\nu},\,q_2^{\mu}p^{\nu},\,p^{\mu}q_2^{\nu},\,q_1^{\mu}p^{\nu}, \\
  \epsilon_{\mu\nu\rho\sigma}p^{\rho}q_1^{\sigma},\,
  \epsilon_{\mu\nu\rho\sigma}p^{\rho}q_2^{\sigma},\,
  \epsilon_{\mu\nu\rho\sigma}q_1^{\rho}q_2^{\sigma}\,.\qquad\qquad\quad
\end{split}
\end{gather}
The antisymmetric tensors are parity-violating.  Electromagnetic gauge
invariance implies that
\begin{align}
q_1^{\mu}T_{\mu\nu} = T_{\mu\nu}\,q_2^{\nu} = 0\,.
\end{align}
From the vanishing of the six coefficients of the linearly independent
vectors, one can deduce five linearly independent conditions. Hence
the most general spin-averaged, gauge-invariant, and parity-conserving
Compton amplitude has five independent form factors:
\begin{align}
  \label{Vmunu}
  \begin{split}
  T^{\mu\nu} &= V_1\left(g^{\mu\nu} - \frac{q_1^{\mu}q_1^{\nu}}{q_1^2} 
    - \frac{q_2^{\mu}q_2^{\nu}}{q_2^2}
    + q_1^{\mu}q_2^{\nu}\frac{(q_1.q_2)}{q_1^2q_2^2} \right) \\
   &+ V_2\left( p^{\mu} - q_1^{\mu}\frac{(p.q_1)}{q_1^2}\right)
   \left(p^{\nu} - q_2^{\nu}\frac{(p.q_2)}{q_2^2}\right) \\
   &+ V_3\left(q_2^{\mu} - q_1^{\mu}\frac{(q_1.q_2)}{q_1^2}\right)
   \left(q_1^{\nu} - q_2^{\nu}\frac{(q_1.q_2)}{q_2^2}\right) \\
   &+ V_4\left( p^{\mu} - q_1^{\mu}\frac{(p.q_1)}{q_1^2}\right)
   \left(q_1^{\nu} - q_2^{\nu}\frac{(q_1.q_2)}{q_2^2}\right) \\
   &+ V_5\left(q_2^{\mu} - q_1^{\mu}\frac{(q_1.q_2)}{q_1^2}\right)
   \left(p^{\nu} - q_2^{\nu}\frac{(p.q_2)}{q_2^2}\right)\,.
  \end{split}
\end{align}
The form factors $V_1$, $V_2$, $V_3$, $V_4$ and $V_5$ can be readily
identified as the coefficients of the tensors $g^{\mu\nu}$,
$p^{\mu}p^{\nu}$, $q_1^{\nu}q_2^{\mu}$, $p^{\mu}q_1^{\nu}$ and
$p^{\nu}q_2^{\mu}$ respectively. They are in general functions of the
six independent scalar invariants. The off-shell Compton form factors
are not directly measurable. The Compton form factors of on-shell
virtual Compton amplitudes, defined by the conditions,
\begin{align}
 p_1^2 = p_2^2 = -M_H^2\,,
\end{align}
depend only on four independent scalar invariants. The gauge-invariant
tensors in \eqref{Vmunu} will be denoted respectively as
$\V_i^{\mu\nu}(p,q_1,q_2)\,,\ i=1,\cdots 5$.

We shall be interested more particularly in two kinematical regimes,
according to whether one or two photons are real. In
electrophotoproduction the outgoing photon is real, $q_2^2=0$, and
thus transversely polarized. One can contract the Compton amplitude
with the polarization $\epsilon_2$, set $\epsilon_2\cdot q_2=0$ and
still impose a gauge condition on the outgoing photon, e.g.
$\epsilon_2\cdot p = 0$, by choosing the Coulomb gauge
$\epsilon_2^0=0$ and the frame $\v{p}=\b{0}$. Then the contracted
amplitude becomes,
\begin{align}
\label{Vmu}
\begin{split}
A^{\mu}_{\text{VCS}} &= T^{\mu\nu}\,\epsilon_{2\nu}^{\star} =
   V_1\left(\epsilon_2^{*\mu} - 
     \frac{q_1^{\mu}}{q_1^2}(\v{\epsilon_2^*}\cdot\v{q_1})\right)
   + V_3\left(q_2^{\mu} - q_1^{\mu}\frac{(q_1.q_2)}{q_1^2}\right)
    (\v{\epsilon_2^*}\cdot\v{q_1}) \\
   &+ V_4\left( p^{\mu} - q_1^{\mu}\frac{(p.q_1)}{q_1^2}\right)
    (\v{\epsilon_2^*}\cdot\v{q_1}) \,.
\end{split}
\end{align}
Therefore there are only three independent form factors when the
outgoing photon is real.

Finally if the ingoing photon is also real, $q_1^2=0$, one can
contract the amplitude $A^{\mu}_{\text{VCS}}$ with the polarization
$\epsilon_1$ and impose similarly the conditions $\epsilon_1\cdot q_1
= \epsilon_1\cdot p = 0$.  Hence the real Compton amplitude has in
general two independent form factors,
\begin{align}
  \label{RCS}
  A_{\text{RCS}} = \epsilon_1^{\mu}\,T_{\mu\nu}\,\epsilon_2^{\star\nu} 
  = V_1\,\v{\epsilon_1}\cdot\v{\epsilon_2}^{\star}
  + V_3\,(\v{\epsilon_1}\cdot \v{q_2})(\v{\epsilon_2}^{\star}\cdot\v{q_1}) \,.
\end{align}

\section{The soft-wall model}
\label{MODEL}

The AdS/CFT correspondence is based upon the fact that the isometry
group of the five-dimensional (5D) anti-de Sitter space is the same as
the four-dimensional conformal group $SO(4,2)$. In Poincar\'e coordinates,
the AdS$_5$ metric reads
\begin{align}
  \label{ADS}
  ds^2 &= \frac{R^2}{z^2}\left(\eta_{\mu\nu}dx^{\mu}dx^{\nu}+dz^2\right)\,,
  \qquad \sqrt{-g} = \frac{R^5}{z^5}\,,
\end{align}
where $\eta_{\mu\nu}\equiv(-1,1,1,1)$ is the four-dimensional
Minkowski metric.  We shall set the curvature radius $R$ to 1 from now on.

Following \cite{PS03} we introduce a massless 5D vector field
$A_m(x,z)$ with a $U(1)$ gauge invariance, which is dual to the
electromagnetic current. The free field $A_m(x,z)$ must satisfy the
Maxwell equations in the bulk (with no dilaton coupling, which would
break the conformal invariance of the electromagnetic field),
\begin{align}
\label{MAXWELL}
    (\nabla_m F)^{mn} = \frac{1}{\sqrt{-g}}\,
    \p_m\left(\sqrt{-g}F^{mn}\right) = 0\,.
\end{align}
Choosing the linear gauge fixing condition,
\begin{align}
  \p^{\mu}A_{\mu} + z\partial_z\left(z^{-1}A_z\right) = 0\,,
\end{align}
the general plane-wave solution of \eqref{MAXWELL} in the space-like
region, $Q^2=q\cdot q>0$, reads
\begin{align}
\begin{split}
  A_{\mu}(x,z) &= \epsilon_{\mu}e^{iq\cdot x}QzK_1(Qz)\,,\\
  A_z(x,z) &= -i\frac{(\epsilon\cdot q)}{q^2}e^{iq\cdot x}
  \p_z\left(QzK_1(Qz)\right)\,,
\end{split}
\end{align}
where $\epsilon_{\mu}$ is a polarization vector.  The boundary
condition is chosen such that the solution becomes a plane-wave on the
Minkowski slice at $z=0$,
\begin{align}
  \label{BC}
  \lim_{z\rightarrow 0} A_{\mu}(x,z) = \epsilon_{\mu}\,e^{iq\cdot x}\,.
\end{align}
The boundary condition in the timelike region, $q^2<0$, can be
obtained by analytic continuation in the $q^2$ variable. A crucial
property of the boundary condition \eqref{BC} is that the vector field
$A_{\mu}(x,z)$ is in fact a constant plane-wave throughout the bulk of
the AdS space when $q^2=0$.

We introduce a massive 5D scalar field $\Phi(x,z)$ which will be the
dual of an operator which creates the spinless target.  Following
\cite{KKSS} the bulk scalar field is coupled to a background dilaton
field $\chi(z)$ which deforms the AdS$_5$ metric. The action which
describes the propagation of $\Phi$ in this background reads
\begin{align}
  S_{\Phi} = \frac{1}{2}\int
  d^4x\,dz\sqrt{-g}e^{-\chi}\left(g^{ij}\partial_i\Phi\partial_j\Phi +
    m^2_S\Phi^2\right)\,,
\end{align}
where $g$ is the AdS$_5$ metric. The classical field equation reads
\begin{align}
  \label{laplace}
  \Delta_g\Phi \equiv \frac{e^{\chi}}{\sqrt{-g}}\partial_i
  \left(e^{-\chi}\sqrt{-g}g^{ij}\p_j\Phi\right) 
  = m_S^2\Phi\,.
\end{align}
In Poincar\'e coordinates, the Laplacian equation becomes
\begin{align}
  \label{scalar}
  z^2\square\Phi + z^5e^{\chi}\partial_z
  \left(z^{-3}e^{-\chi}\partial_z\Phi\right) 
  = m_S^2\Phi\,.
\end{align}
Looking for a solution that is a plane-wave in Minkowski space and setting
\begin{align}
  \Phi(x,z) = e^{ip\cdot x}e^{\chi(z)/2}z^{3/2}\psi(z) \equiv 
  e^{ip\cdot x}\,\h{\Phi}(z) \,,  
\end{align}
the Laplacian equation is transformed into a Schr\"odinger-like equation
\begin{gather}
  \frac{d^2\psi}{dz^2} - V(z)\psi = p^2\psi\,, \\
  V(z) = \frac{m_S^2+15/4}{z^2} + \frac{3}{2z}\partial_z\chi
  + \frac{1}{4}(\partial_z\chi)^2 - \frac{1}{2}\partial_z^2\chi\,.
\end{gather}
If the potential $V(z)$ has the right properties, and with appropriate
boundary conditions, this equation has a complete
set of solutions $\{\psi_n(z)\,,\ n\in\NB\}$ which form an
orthonormal basis of the Hilbert space $\H$ defined by the inner
product
\begin{gather}
  \langle \phi|\psi\rangle_{\H} = \int_0^{\infty}dz\,\phi^{\star}(z)\psi(z) \,.
\end{gather}
We shall assume that the dilaton background is such that $\H$ is
well-defined.  Otherwise we let the dilaton profile unspecified to be
as generic as possible, except that conformal invariance in the
ultraviolet requires that $\chi(z)\rightarrow 0$ for $z\rightarrow 0$.

In terms of the plane-wave solutions of the Laplacian equation
\eqref{scalar}, with the corresponding boundary conditions, the
completeness relation takes the form
\begin{align}
 \delta(z-z') = \sum_n z'^{-3/2}e^{-\chi(z')/2}\,\h{\Phi}_n^{\star}(z')\,
 \h{\Phi}_n(z)\,z^{-3/2}e^{-\chi(z)/2}\,.
\end{align}
Therefore the set of classical solutions $\{\h{\Phi}_n(z)\,,\
n\in\NB\}$ form a complete orthonormal basis of the Hilbert space
$\H_S$ spanned by the solutions of the Laplacian equation and defined
by the inner product,
\begin{align}
\label{HS}
 \left\langle\h{\Phi}\right|\left.\h{\Phi}'\right\rangle_{\H_S} = 
 \int_0^{\infty}dz\,z^{-3}e^{-\chi(z)}\,
 \h{\Phi}^{\star}(z)\,\h{\Phi}'(z) \,.
\end{align}
Each $\h{\Phi}_n$ is a normalized eigenfunction, with appropriate
boundary conditions, of the operator
\begin{align}
  \label{H}
    \h{H}_S\,\h{\Phi}_n = \left(z^3\,e^{\chi}\partial_z
  \left(z^{-3}e^{-\chi}\partial_z\right) - m_S^2z^{-2}\right)\h{\Phi}_n
= -m_n^2\h{\Phi}_n\,.
\end{align}

The scalar Green function $G(x,z;x',z')$ is defined by the
inhomogeneous equation
\begin{align}
  \label{Green}
   \left(\Delta_g - m_S^2\right) G(x,z;x',z') = \frac{e^{\chi}}{\sqrt{-g}}
 \delta^{4}(x-x')\delta(z-z')\,.
\end{align}
Its four-dimensional Fourier transform $\h{G}(z,z';p)$
\begin{align}
  \label{FT}
  G(x,z;x',z') = 
  \frac{1}{(2\pi)^4}\int_{-\infty}^{+\infty}d^4p
  \,e^{ip(x'-x)}\h{G}(z,z';p) 
\end{align}
satisfies the equation
\begin{align}
  \label{green}
  \h{H}_S\,\h{G} = p^2\h{G} + z^3e^{\chi}\,\delta(z-z') \,.
\end{align}
$\h{G}$ has an expansion in terms of the normalized
eigenfunctions satisfying the same boundary conditions,
\begin{align}
  \label{expansion}
  \h{G}(z,z';p) &= -\sum_{n=0}^{\infty} 
  \frac{\h{\Phi}^{\star}_n(z)\h{\Phi}_n(z')}{p^2+m^2_n-i\epsilon} \,.
\end{align}
The handling of the singularities at $p^2=-m_n^2$ is done with the
standard Feynman prescription.

In order to complete the definition of the model we still need to
specify the interaction between the bulk scalar field and the bulk
$U(1)$ field. Since we are interested in describing the
electromagnetic interactions of a charged spinless hadron, we shall
take a $U(1)$ covariant coupling, \DS{D_n\Phi = \p_n\Phi -ie A_n\Phi}.
Hence we shall consider the full anti-de Sitter action,
\begin{equation}
\label{AdS}
S_{\text{AdS}}[\Phi,\Phi^*,A^m]=\int d^4xdz\ \sqrt{-g}
\left(-\frac{1}{4}F^{mn}F_{mn} 
  + e^{-\chi}\left((D^m\Phi)^*D_m\Phi+m_S^2\Phi^*\Phi\right)\right)\ .
\end{equation}

\section{Calculation of four-point functions in AdS/QCD}
\label{4PT}

The gauge/gravity correspondence relates generating functions in a
strongly-coupled gauge theory to the classical supergravity
partition function in the following way:
\begin{align}
\label{corres}
\begin{split}
Z_{CFT}(c,\bar{c},n+\bar{n}) &= \left\langle\exp\left(\int d^4x\ 
(n_\mu+\bar{n}_\mu)J^\mu+\bar{c}\,O+c\,O^\dagger\right)\right\rangle_{CFT} \\
&= \exp\left(-S^{cl}_{\text{AdS}}
[\Phi(c),\Phi^*(\bar{c}),A^m(n_\mu+\bar{n}_\mu)]
\right) \,,
\end{split}
\end{align}
where the 4-dimensional sources for the CFT appear as boundary
conditions for the 5-dimensional classical supergravity fields.
Correlation functions of CFT operators can be obtained by expanding
\eqref{corres} to linear-order with respect to the sources. We shall
use the prescription \eqref{corres} as a recipe for the AdS/QCD model.
The correlation function we are interested in can be
obtained from the coefficient of $\bar{c}n_\mu\bar{n}_\nu c.$

In the contracted Compton amplitude, $\epsilon_{1\mu}
T^{\mu\nu}\epsilon_{2\nu}^*,$ the QCD operators are coupled to
asymptotic states, therefore these will serve as boundary conditions
for the free bulk fields $\Phi^{(0)},$ $\Phi^{*(0)},$ and
$A_\mu^{(0)}.$ After we express $S^{\text{cl}}_{\text{AdS}}$ in terms
of free fields, it will be easy to read off the
$\bar{c}n_\mu\bar{n}_\nu c$ coefficient.

With the notations of the previous section, the equations of motion
for the interacting classical bulk fields read
\begin{align}
\frac{1}{\sqrt{-g}}\partial_n\left(\sqrt{-g}F^{mn}\right)
&= ie\,e^{-\chi}\left(\Phi^* D^m\Phi-(D^m\Phi^*)\Phi\right) \,, \\
(\Delta_g-m_S^2)\Phi &= V(A)\Phi=ieV_1(A)\Phi+e^2V_2(A)\Phi \,, \\
(\Delta_g-m_S^2)\Phi^* &= \o{V}(A)\Phi^*=-ieV_1(A)\Phi^*+e^2V_2(A)\Phi^* \,,
\end{align}
where the linear operators $V_1(A)$ and $V_2(A)$ act on the right and read
\begin{align}
V_1(A)&=\frac{e^\chi}{\sqrt{-g}}\,
\partial_m\left(\sqrt{-g}e^{-\chi}A^m\right)+2A^m \partial_m \,,
\\ V_2(A)&=A^m A_m \,.
\end{align}
The function $V$ can be also used to write the interaction term in
$S_{\text{AdS}}:$
\begin{align}
S_{int}&= \int d^4xdz\ \sqrt{-g}e^{-\chi}
\left(ieA^m(\Phi^*\partial_m\Phi-\Phi\partial_m\Phi^*)
+e^2A^mA_m\Phi^*\Phi\right) \,,
\nonumber\\
&= \frac{1}{2}\int d^4xdz\ \sqrt{-g}e^{-\chi}
\left(\Phi^*V(A)\Phi+\Phi\o{V}(A)\Phi^*\right) \,.
\end{align}
The solutions for $\Phi$ and $\Phi^*$ can be written as
\begin{align}
\Phi^{(*)}(y)&=\Phi^{(*)}_{(0)}(y)+\int dy'\sqrt{-g'}e^{-\chi(z')}G(y;y')\,
\overset{(-)}{V}\left(A(y')\right)\Phi^{(*)}(y') \,,
\end{align}
where the free bulk fields $\Phi_{(0)}$, $\Phi^*_{(0)}$ and the Green
function $G$ are respectively solutions of \eqref{laplace} and
\eqref{green}. We use the shorthand notations $y=(x,z)$ and
$dy=d^4xdz$. We can now write $\Phi^*V(A)\Phi$ in terms of the free
fields:
\begin{align}
\label{phivphi}
\begin{split}
\Phi^*V(A)\Phi &= \left(\Phi^*_{(0)}(y) -
ie\int dy'\sqrt{-g'}e^{-\chi(y')}\,G(y;y')
\,V_1\left(A_{(0)}(y')\right)\Phi^*_{(0)}(y')\right) \\
&\times\left(ieV_1\left(A_{(0)}(y)\right) + e^2 V_1\left(A_{(1)}(y)\right)
+e^2 V_2\left(A_{(0)}(y)\right)\right) \\
&\times\left(\Phi_{(0)}(y) + ie\int dy''\sqrt{-g''}e^{-\chi(y'')}\,G(y;y'')
\,V_1\left(A_{(0)}(y'')\right)\Phi_{(0)}(y'')\right) + {\cal O}(e^3) \,,
\end{split}
\end{align}
and similarly for $\Phi\o{V}(A)\Phi^*.$ In $S^{cl}_{int},$ the
contribution involving $A_{(0)} A_{(0)} \Phi^*_{(0)}\Phi_{(0)}$
appears at order $e^2:$
\begin{align}
\begin{split}
S^{cl}_{int} &= \int dy\,\sqrt{-g}e^{-\chi}
\left(ieA^m(\Phi^*\partial_m\Phi-\Phi\partial_m\Phi^*)
+e^2A^m A_m\Phi^*\Phi\right) \\
&+ e^2\int dydy'\,\sqrt{-g}e^{-\chi(y)}\sqrt{-g'}e^{-\chi(y')} A^m(y)
\left\{\left(G(y;y')\partial_m\Phi^*(y)\right.\right.
\\ &\left.\left.\qquad\qquad-\Phi^*(y)\partial_m G(y;y')\right) 
V_1(A(y'))\Phi(y')+\left(\Phi\leftrightarrow\Phi^*\right)\right\}
+{\cal O}(e^3) \,,
\end{split}
\end{align}
where we have dropped the ${}_{(0)}$ notation for clarity, and we are
now dealing only with free fields. Note that the term
$V_1(A_{(1)})$ in (\ref{phivphi}) contributes also at order
$e^2$ but we have dropped it since it does not contribute to
$T^{\mu\nu},$ but rather to a $\langle\Phi^{*2}\Phi^2\rangle$
correlator. Finally, after integrating by parts over $y'$ one writes:
\begin{align}
S^{cl}_{int} &= ie\int dy\,\sqrt{-g}e^{-\chi}
A^m(\Phi^*\partial_m\Phi-\Phi\partial_m\Phi^*)
+ e^2\int dy\,\sqrt{-g}e^{-\chi} A^m A_m\Phi^*\Phi \nonumber\\
&+ e^2\int dydy'\,\sqrt{-g}e^{-\chi} \sqrt{-g'}e^{-\chi'}
A^m(y)A^n(y') \nonumber\\
&\qquad\quad\left\{
\Bigl(\Phi^*(y)\partial_m-(\partial_m\Phi^*(y))\Bigl)
\Big(\Phi(y')\partial'_n-(\partial'_n\Phi(y'))\Bigl)\right.\nonumber\\
&\qquad\quad\left.+\Bigl(\Phi(y)\partial_m-(\partial_m\Phi(y))\Bigl)
\Bigl(\Phi^*(y')\partial'_n-(\partial'_n\Phi^*(y'))\Bigl)
\right\}G(y;y')\,.
\end{align}

In this expression, the boundary conditions at $z=0$ of the classical
fields $A^m,$ $\Phi,$ and $\Phi^*$ are respectively
$n_{\mu}+\bar{n}_{\mu},$ $c,$ and $\bar{c}.$ These enter in a linear
way in the fields, therefore the $\bar{c}n_\mu\bar{n}_\nu c$
coefficient in the expansion of \eqref{corres} is simply obtained.
After contractions, one can write
\begin{align}
  \epsilon_\mu T^{\mu\nu}\epsilon^*_\nu
  &=2 e^2\int dy\,\sqrt{-g}e^{-\chi}\ 
  A^m(y)A^*_m(y)\Phi^*(y)\Phi(y)\nonumber\\
  &+e^2\int dydy'\,\sqrt{-g}e^{-\chi} \sqrt{-g'}e^{-\chi'} 
  \left(A^m(y)A^{*n}(y')+A^{*m}(y)A^n(y')\right)
  \nonumber\\&\quad\times
  \Big(\Phi(y)\partial_m-(\partial_m\Phi(y))\Big)\ 
  \Big(\Phi^*(y')\partial'_n-(\partial'_n\Phi^*(y'))\Big)
  G(y;y')\,,
\label{fullTmunu}
\end{align}
where the fields $\Phi(x,z)$ and $A(x,z)$ are the plane-wave solutions
defined in section \ref{MODEL}.

In a diagrammatic representation, the first contribution in
(\ref{fullTmunu}) is a contact interaction, while the second
contribution contains an $s$-channel diagram (with the term
$A^m(x,z)A^{*n}(x',z')$) and a $u$-channel diagram (with the term
$A^{*m}(x,z)A^{n}(x',z')$) \cite{GX09}. This is not surprising, since
it is well-known that taking the classical limit of a quantum field
theory is equivalent to keeping only tree diagrams in the perturbative
expansion.

\section{On-shell Compton amplitude}
\label{SCALAR}

The $s$-channel contribution in eq.\,\eqref{fullTmunu} can be written as
\begin{align}
  \begin{split}
  \epsilon^{\mu}T_{\mu\nu}^s\epsilon^{\star\nu} &=
  (ie)^2\int d^4x\,d^4x'\,dz\,dz'\,z^{-3}e^{-\chi(z)}A_k(x,z)
  A^{\star}_l(x',z')z'^{-3}e^{-\chi(z')} \\
  &\qquad\qquad\quad\times
  \left(\Phi_{\text{in}}(x,z)\op{\p^k}_{(x,z)}G(x,z;x',z')
  \op{\p^l}_{(x',z')}\Phi^{\star}_{\text{out}}(x',z')\right) \,,
  \end{split}
\end{align}
where the initial and final wave-functions of the bulk scalar fields,
$\Phi_{\text{in}}$ and $\Phi_{\text{out}}$, are normalized plane-wave
solutions of the Laplacian equation \eqref{scalar}, whereas the bulk
vector field $A_k$ is a normalized plane-wave solution of the Maxwell
equations. The Green function $G(x,z;x',z')$ is defined by
eqs.\,\eqref{FT} and \eqref{green}. We shall introduce the shorthand
notations,
\begin{align}
  \label{notations}
  \begin{split}
  &\Phi_{\text{in}}(x,z) = e^{ip_1\cdot x}\h{\Phi}_i(z)\,,\quad
   \Phi_{\text{out}}(x',z') = e^{ip_2\cdot x'}\h{\Phi}_f(z')\,, \\
  &A_{\mu}(x,z) = \epsilon_{\mu}e^{iq_1\cdot x}A_1(z)\,,\quad
   A_z(x,z) = -i\frac{\epsilon\cdot q_1}{q_1^2}e^{iq_1\cdot x}\p_zA_1(z)\,, \\
  &A_{\nu}(x',z') = \epsilon_{\nu}e^{iq_2\cdot x'}A_2(z')\,,\quad
   A_z'(x',z') = -i\frac{\epsilon\cdot q_2}{q_2^2}
   e^{iq_2\cdot x'}\p_{z'}A_2(z')\,,\\
  &A_1(z) = Q_1z\,K_1(Q_1z) \,,\quad A_2(z) = Q_2z\,K_1(Q_2z) \,.
\end{split}
\end{align}
A straightforward calculation yields,
\begin{align}
  \begin{split}
T_{\mu\nu}^s &= (2\pi)^4\delta^{(4)}(p_1+q_1-p_2-q_2)\,e^2\times\biggl(
 (p_1+k)_{\mu}(p_2+k)_{\nu}\,\F_1\left(q^2_1,q_2^2,s\right) \\
&\quad
- \frac{(p_1+k)_{\mu}q_{2\nu}}{q_2^2}\,\F_2\left(q^2_1,q_2^2,s\right) 
+ \frac{q_{1\mu}(p_2+k)_{\nu}}{q_1^2}\,\F_3\left(q^2_1,q_2^2,s\right)
- \frac{q_{1\mu}q_{2\nu}}{q_1^2q_2^2}\,\F_4\left(q^2_1,q_2^2,s\right)
\biggr)\,.
  \end{split}
\end{align}
with $k=p_1+q_1=p_2+q_2$, $s=k^2$, and
\begin{align}
  \begin{split}
  \F_1\left(q^2_1,q_2^2,k^2\right) &=  
  \iint dz\,dz'\,\,z^{-3}e^{-\chi(z)}A_1(z)\h{\Phi}_i(z)\h{G}(z,z',k^2)
  \h{\Phi}_f^{\star}(z')A_2^{\star}(z')z'^{-3}e^{-\chi(z')}\,, \\
  \F_2\left(q^2_1,q_2^2,k^2\right) &=  
  \iint dz\,dz'\,\,z^{-3}e^{-\chi(z)}A_1(z)\h{\Phi}_i(z)\left(\h{G}(z,z',k^2)
    \op{\p}_{z'}\h{\Phi}_f^{\star}(z')\right)
  \p_{z'}A_2^{\star}(z')z'^{-3}e^{-\chi(z')}\,,\\
  \F_3\left(q^2_1,q_2^2,k^2\right) &=  
  \iint dz\,dz'\,\,z^{-3}e^{-\chi(z)}\p_zA_1(z)
  \left(\h{\Phi}_i(z)\op{\p}_z\h{G}(z,z',k^2)\right)
  \h{\Phi}_f^{\star}(z')A_2^{\star}(z')z'^{-3}e^{-\chi(z')}\,, \\
  \F_4\left(q^2_1,q_2^2,k^2\right) &=  
  \iint dz\,dz'\,\,z^{-3}e^{-\chi(z)}\p_zA_1(z)
  \left(\h{\Phi}_i(z)\op{\p}_z\h{G}(z,z',k^2)
    \op{\p}_{z'}\h{\Phi}_f^{\star}(z')\right)
  \p_{z'}A_2^{\star}(z')z'^{-3}e^{-\chi(z')}\,.
  \end{split}
\end{align}
The $u$-channel amplitude is related to the $s$-channel amplitude
according to the crossing symmetry by interchanging $\mu\lr\nu$, 
$q_1\lr -q_2$,\ $\epsilon\lr \epsilon^{\star}$ and
$A_1\lr A_2^{\star}$. Hence it reads
\begin{align}
\begin{split}
  T_{\mu\nu}^u &=  (2\pi)^4\delta^{(4)}(p_1+q_1-p_2-q_2)\,e^2 
  \times\biggl( (p_2+k')_{\mu}(p_1+k')_{\nu}\,\F_1(q_2^2,q_1^2,u) \\
  &\quad 
  + \frac{q_{1\mu}(p_1+k')_{\nu}}{q_1^2}\,\F_2(q_2^2,q_1^2,u)
  - \frac{(p_2+k')_{\mu}q_{2\nu}}{q_2^2}\,\F_3(q_2^2,q_1^2,u)
  - \frac{q_{1\mu}q_{2\nu}}{q_1^2q_2^2}\,\F_4(q_2^2,q_1^2,u) \biggr)\,,
  \end{split}
\end{align}
with $k' = p_1-q_2 = p_2-q_1$ and $u=k'^2$.

Integrating by parts the partial derivative of $A_1$,
\begin{align}
  \F_3(q_1^2,q_2^2,k^2) &= -\iint dz\,dz'\,\,z'^{-3}e^{-\chi(z')}
  A^{\star}_2(z')\h{\Phi}_f^{\star}(z')A_1(z)\p_{z}
  \left(z^{-3}e^{-\chi(z)}\left(\h{\Phi}_i(z)\op{\p}_z\h{G}(z,z',k^2)\right)
  \right)\,,
\end{align}
and using equations \eqref{scalar} and \eqref{green}, one gets
\begin{align}
\nonumber
 &\p_{z}\left(z^{-3}e^{-\chi(z)}\left(\h{\Phi}_i(z)\op{\p}_z\h{G}(z,z',k^2) 
   \right)\right) = \\
\nonumber
 &\h{\Phi}_i(z)\p_{z}\left(z^{-3}e^{-\chi(z)}\p_z\h{G}(z,z',k^2)\right) 
 -\p_{z}\left(z^{-3}e^{-\chi(z)}\p_z\h{\Phi}_i(z)\right)\h{G}(z,z',k^2) = \\
\label{rel}
 & z^{-3}e^{-\chi}\left(k^2-p_1^2\right)\h{\Phi}_i(z)\h{G}(z,z',k^2)
     + \delta(z-z')\h{\Phi}_i(z) \,.
\end{align}
Hence
\begin{align}
  \F_3(q_1^2,q_2^2,k^2) = (p_1^2-k^2)\F_1(q_1^2,q_2^2,k^2) - 
  \int dz\, z^{-3}e^{-\chi(z)}A_1(z)A_2^{\star}(z)
  \h{\Phi}_i(z)\h{\Phi}_f^{\star}(z)\,.
\end{align}
Similarly,
\begin{align}
  \F_2(q_1^2,q_2^2,k^2) &=
   -\iint dz\,dz'\,z^{-3}e^{-\chi(z)}A_1(z)\h{\Phi}_i(z)A_2^{\star}(z')
  \p_{z'}\left(z'^{-3}e^{-\chi(z')}\left(\h{G}(z,z',k^2)
      \op{\p}_{z'}\h{\Phi}_f^{\star}(z')\right)\right)\,, \nonumber\\
  &=  (k^2-p_2^2)\F_1(q_1^2,q_2^2,k^2) + \int dz\, z^{-3}e^{-\chi(z)}
  A_1(z)A_2^{\star}(z)\h{\Phi}_i(z)\h{\Phi}_f^{\star}(z)\,.
\end{align}
The four-point interaction amplitude reads
\begin{align}
  \label{CONTACT}
  \epsilon^{\mu}T^c_{\mu\nu}\epsilon^{\star\nu} &= 
  -2(ie)^2\int d^4x\,dz\,\sqrt{-g}e^{-\chi}
  \h{\Phi}_i(x,z)g^{mn}A_m(x,z)A_n^{\star}(x,z)\h{\Phi}_f^{\star}(x,z)\,, 
  \nonumber\\
  \begin{split}
  T^c_{\mu\nu} &= 2 e^2(2\pi)^4\delta^{(4)}(p_1+q_1-p_2-q_2) \times \\
  &\quad\int dz z^{-3}e^{-\chi}\,
  \h{\Phi}_i(z)\left(g_{\mu\nu}A_1(z)A_2^{\star}(z) + 
    \frac{q_{1\mu}q_{2\nu}}{q_1^2q_2^2}
    \p_zA_1(z)\p_zA_2^{\star}(z)\right)\h{\Phi}_f^{\star}(z) \,.
  \end{split}
\end{align}
Let
\begin{align}
  \C_1(q_1^2,q_2^2) &= \C_1(q_2^2,q_1^2) = \int dz\,z^{-3}e^{-\chi(z)}
  \,A_1(z)\,A_2^{\star}(z)\,\h{\Phi}_i(z)\,\h{\Phi}_f^{\star}(z) \,, \\
  \C_0(q_1^2,q_2^2) &= \C_0(q_2^2,q_1^2) =
  \int dz\, z^{-3}e^{-\chi(z)}\,\p_zA_1(z)\,\p_zA_2^{\star}(z)
  \,\h{\Phi}_i(z)\,\h{\Phi}_f^{\star}(z) \,.
\end{align}
Setting $p=p_1+p_2$, the total amplitude can be written as
\begin{align*}
  T_{\mu\nu} &= e^2(2\pi)^4\delta^{(4)}(p_1+q_1-p_2-q_2) \times \biggl(\\
  &\quad 
  \left(\F_1\left(q_1^2,q_2^2,s\right) +
    \F_1\left(q_2^2,q_1^2,u\right)\right)\times
  \biggl(\left(p_{\mu}p_{\nu} - \frac{p\cdot q_2}{q_2^2}p_{\mu}q_{2\nu} 
    - \frac{p\cdot q_1}{q_1^2}p_{\nu}q_{1\mu}\right) + \\
  & \qquad\qquad\qquad\qquad\qquad\qquad\qquad\qquad
  \left(q_{1\nu}q_{2\mu} - \frac{q_1\cdot q_2}{q_1^2}q_{1\mu}q_{1\nu}
  - \frac{q_1\cdot q_2}{q_2^2}q_{2\mu}q_{2\nu}\right)\biggr) \\
  &+ \left(\F_1\left(q_1^2,q_2^2,s\right)-\F_1\left(q_2^2,q_1^2,u\right)\right)
  \times\biggl(
  \left(p_{\mu}q_{1\nu} - \frac{q_1\cdot q_2}{q_2^2}p_{\mu}q_{2\nu} 
  - \frac{p\cdot q_1}{q_1^2} q_{1\mu}q_{1\nu}\right) + \\
  & \qquad\qquad\qquad\qquad\qquad\qquad\qquad\qquad
  \left(p_{\nu}q_{2\mu} - \frac{q_1\cdot q_2}{q_1^2}p_{\nu}q_{1\mu} -
  \frac{p\cdot q_2}{q_2^2}q_{2\mu}q_{2\nu}\right)\biggr) \\
  &\ + 2\C_1\left(q_1^2,q_2^2\right)\left(g_{\mu\nu} 
    - \frac{q_{1\mu}q_{1\nu}}{q_1^2}
    - \frac{q_{2\mu}q_{2\nu}}{q_2^2}\right)
    + \left(2\C_0(q_1^2,q_2^2) - \F_4\left(q_1^2,q_2^2,s\right) - 
    \F_4\left(q_2^2,q_1^2,u\right)\right)\frac{q_{1\mu}q_{2\nu}}{q_1^2q_2^2}
  \biggr) \,.  
\end{align*}
Comparing with the gauge-invariant tensor basis \eqref{Vmunu}, gauge
invariance holds true if, and only if,
\begin{align}
  \begin{split}
  2\C_0(q_1^2,q_2^2) &=  \F_4\left(q_1^2,q_2^2,s\right)
  +\F_4\left(q_2^2,q_1^2,u\right)  \\
  &+ \left(\F_1\left(q_1^2,q_2^2,s\right)+\F_1\left(q_2^2,q_1^2,u\right)\right)
  \left((p\cdot q_1)(p\cdot q_2)+(q_1\cdot q_2)^2\right) \\
  &+ \left(\F_1\left(q_1^2,q_2^2,s\right)-\F_1\left(q_2^2,q_1^2,u\right)\right)
  \left(p\cdot q_1 + p\cdot q_2\right)(q_1\cdot q_2) \\
  &+ 2\C_1(q_1^2,q_2^2)(q_1\cdot q_2)\,.
  \end{split}
\end{align}
Expanding the bidirectional derivatives, the form factor $\F_4$ reads
\begin{align}
\begin{split}
  \F_4 &= \iint dz\,dz'\,z^{-3}e^{-\chi(z)}\,\p_zA_1(z)\times \\
  &\qquad\qquad\biggl(
  \left(\h{\Phi}_i(z)\,\p_z\h{G} - \h{G}\,\p_z\h{\Phi}_i(z)\right)
  \p_{z'}\h{\Phi}_f^{\star}(z')
  - \p_{z'}\left(\h{\Phi}_i(z)\,\p_z\h{G} - \p_z\h{\Phi}_i(z)\,\h{G}\right)
  \h{\Phi}_f^{\star}(z') \biggr) \\
  &\qquad\quad\times \p_{z'}A_2^{\star}(z')z'^{-3}e^{-\chi(z')} \,.
  \end{split}
\end{align}
We first integrate by parts over $\p_zA_1(z)$ and use equation
\eqref{rel} in the $s$-channel,
\begin{align}
  \nonumber
  \begin{split}
  \F_4(s) &= -\iint dz\,dz'\,A_1(z)\times \\
  &\qquad\qquad\biggl(
  \left(z^{-3}e^{-\chi(z)}\left(s-p_1^2\right)\h{\Phi}_i(z)\h{G}(z,z',s)
     + \delta(z-z')\,\h{\Phi}_i(z)\right)\p_{z'}\h{\Phi}_f^{\star}(z') \\
  &\qquad\qquad 
  -\p_{z'}\left(z^{-3}e^{-\chi}\left(s-p_1^2\right)\h{\Phi}_i(z)\h{G}(z,z',s)
     + \delta(z-z')\,\h{\Phi}_i(z)\right)\h{\Phi}_f^{\star}(z') \biggr) \\
  &\qquad\quad\times \p_{z'}A_2^{\star}(z')z'^{-3}e^{-\chi(z')} \,,
  \end{split}
  \\
  \begin{split}
  &= (p_1^2-s)\iint dz\,dz'\,z^{-3}e^{-\chi(z)}\,A_1(z)\,\h{\Phi}_i(z)\times \\
  & \qquad\qquad\qquad\quad\left(\h{G}\,\p_{z'}\h{\Phi}_f^{\star}(z') - 
    \h{\Phi}_f^{\star}(z')\,\p_{z'}\h{G}\right)
  \p_{z'}A_2^{\star}(z')\,z'^{-3}e^{-\chi(z')} \\
  &\quad-
  \int dz\,z^{-3}e^{-\chi(z)}\,A_1(z)\,\h{\Phi}_i(z)\,\p_z\h{\Phi}_f^{\star}(z)
  \,\p_zA_2^{\star}(z) \\
  &\quad- \int dz\,A_1(z)\,\h{\Phi}_i(z)\,\p_z
  \left(\h{\Phi}_f^{\star}(z)\,\p_zA_2^{\star}(z)\,z^{-3}e^{-\chi(z)}\right)\,.
  \end{split}
\end{align}
We now integrate the first term by parts over $\p_{z'}A^{\star}_2(z')$,
use again equation \eqref{rel}, and integrate by parts the third term,
\begin{align}
  \begin{split}
  \F_4(s) &= (p_1^2-s)
  \iint dz\,dz'\,z^{-3}e^{-\chi(z)}\,A_1(z)\,\h{\Phi}_i(z)\times
  \\ & \qquad\qquad\qquad\quad
  \left(z'^{-3}e^{-\chi(z')}\left(s-p_2^2\right)\h{G}(z,z',s)
     + \delta(z-z')\right)\h{\Phi}_f^{\star}(z')\,A_2^{\star}(z') \\
  &\quad- 
  \int dz\,z^{-3}e^{-\chi(z)}\,A_1(z)\,\h{\Phi}_i(z)\,\p_z\h{\Phi}_f^{\star}(z)
  \,\p_zA_2^{\star}(z) \\
  &\quad+ \int dz\,z^{-3}e^{-\chi(z)}\,\p_z\left(A_1(z)\,\h{\Phi}_i(z)\right)
  \,\h{\Phi}_f^{\star}(z)\,\p_zA_2^{\star}(z) \,.
  \end{split}
  \end{align}
Hence
\begin{align}
  \label{T4}
  \begin{split}
  \F_4(s) &= -(p_1^2-s)(p_2^2-s)\F_1(s) + (p_1^2-s)\C_1 +\C_0 \\
  &\quad- \int dz\,z^{-3}e^{-\chi(z)}\,A_1(z)
  \left(\h{\Phi}_i(z)\,\op{\p}_z\h{\Phi}_f^{\star}(z)\right)
  \,\p_zA_2^{\star}(z)\,.
  \end{split}
\end{align}
A similar identity holds in the $u$-channel. Noting that
\begin{gather}
  \nonumber
  \begin{split}
  (p\cdot q_1)(p\cdot q_2)+(q_1\cdot q_2)^2 \pm 
  (p\cdot q_1 + p\cdot q_2)(q_1\cdot q_2) = 
  (p\pm q_1)\cdot q_2 \times (p\pm q_2)\cdot q_1 \,,
   \end{split}
  \\
  \begin{split}
  (p+q_2)\cdot q_1\times(p+q_1)\cdot q_2 = (p_1^2-s)(p_2^2-s) \,, \\
  (p-q_2)\cdot q_1\times(p-q_1)\cdot q_2 = (p_1^2-u)(p_2^2-u) \,,
 \end{split}
\end{gather}
gauge invariance is recovered when $p_1^2=p_2^2$. Indeed, then
$\h{\Phi}_i(z)=\h{\Phi}_f(z)$ since they satisfy the same equation, so the
last term in eq.\,\eqref{T4} vanishes.

To summarize, the on-shell Compton amplitude reads, with
$p_1^2=p_2^2=-m^2$,
\begin{align}
  \label{SCA}
  \begin{split}
  T^{\mu\nu} &=  e^2\left(2\C_1\,\V_1^{\mu\nu} 
    + \C_+\left(\V_2^{\mu\nu} + \V_3^{\mu\nu}\right)
    + \C_-\left(\V_4^{\mu\nu} + \V_5^{\mu\nu}\right)\right) \,,
\end{split}
\end{align}
where the tensors $\V_i^{\mu\nu},\ i=1,\cdots 5$, are defined in
eqs.~\eqref{Vmunu} and
\begin{align}
  C_{\pm}(m^2,q_1^2,q_2^2,s,u) = \F_1(m^2,q_1^2,q_2^2,s) \pm 
  \F_1(m^2,q_2^2,q_1^2,u) \,.
\end{align}
The $\delta$ factor expressing energy-momentum conservation is
implicitly understood from now on in all formulas for the Compton
amplitude. The form factors $\F_1(m^2,q_1^2,q_2^2,k^2)$ and
$\C_1(m^2,q_1^2,q_2^2)$ are defined by
\begin{align}
  \label{SVCF}
  \begin{split}
  \F_1\left(m^2,q^2_1,q_2^2,k^2\right) &=  
  \iint dz_1\,dz_2\,\,z_1^{-3}e^{-\chi(z_1)}\,A_1(z_1)\,\h{\Phi}_i(z_1)\,
  \h{G}(z_1,z_2,k^2)
  \,\h{\Phi}_f^{\star}(z_2)\,A_2^{\star}(z_2)\,z_2^{-3}e^{-\chi(z_2)}\,, \\
  \C_1(m^2,q_1^2,q_2^2) &= \int dz\,z^{-3}e^{-\chi(z)}
  \,A_1(z)\,A_2^{\star}(z)\,\h{\Phi}_i(z)\,\h{\Phi}_f^{\star}(z) \,.
  \end{split}
\end{align}
These formulas are valid for any dilaton background $\chi(z)$ on the
5D AdS space which yields a well-defined inner product on the vector
space of solutions of the classical field equations. If $\chi\equiv
0$, our formulas coincide with the Compton amplitude off a dilaton
calculated in \cite{GX09}, once we identify the normalized functions
$\h{\Phi}_i$ and $\h{\Phi}_f$ with the Bessel solutions of the
hard-wall model. We have shown quite generally that the
gauge-invariant structure of the Compton amplitude is similar for
hard-wall and soft-wall models.  Note that we only get three
independent Compton form factors out of a possible five. These generic
properties do not depend upon the special way of breaking the
conformal invariance for extending the AdS/CFT correspondence to QCD,
nor upon detailed relations between Bessel functions or other special
functions. It depends only upon the general structure of the classical
equations satisfied by the bulk scalar fields and their Green
functions in AdS space with a dilaton background.

Moreover we have shown explicitly that the off-shell Compton
amplitudes ($p_1^2 \ne p_2^2$) calculated in these simplest AdS/QCD
models are not gauge-invariant. This result is also not obvious and is
in fact very specific to single scalar intermediate states. For
example, had we allowed for non-minimal couplings with only single
vector intermediate states, we would get gauge-invariant off-shell
Compton amplitudes. One should perhaps emphasize that the non-gauge
invariance of the off-shell four-point Compton amplitude has nothing
to do with the (trivial) non-gauge invariance of the three-point
function when the conformal dimensions of the initial and final scalar
states are different. In our case, the initial, final, and
intermediate scalar states are solutions of the same classical field
equation in AdS space and have the same conformal dimension. The
non-gauge invariance, when the initial and final scalar states are not
the same mass eigenstates in Minkowski space, is due to the particular
form of the contact term \eqref{CONTACT}.

\section{Deeply inelastic scattering}
\label{DIS}

The Compton amplitude \eqref{SCA} has a rather simple structure which
it is enlightening to unravel. Expanding the Green function over the
orthonormal eigenstates, we can write the form factor $\F_1$ in the
doubly spacelike region as
\begin{align}
  \label{IF}
  \F_1\left(m^2,q^2_1,q_2^2,k^2\right) &=  -\sum_{n=0}^{\infty}
  \frac{\Gamma(m^2,m_n^2,q_1^2)\,\Gamma^*(m^2,m_n^2,q_2^2)}
  {k^2+m^2_n-i\epsilon} \,, \\
  \label{IV}
  \Gamma(m^2,m_n^2,Q^2) &= Q\int dz\,z^{-2}e^{-\chi(z)}
  K_1(Qz)\,\h{\Phi}(z)\,\h{\Phi}^*_n(z) \,.
\end{align}
It is immediate to show from the AdS/QCD dictionary that the vertex
function $\Gamma$ is just the unique form factor which parametrizes
the most general matrix element of the conserved electromagnetic
current between two (pseudo)scalar states,
\begin{align}
  \label{FF}
  \langle p_2|J^{\mu}(0)|p_1\rangle = \Gamma(p_1^2,p_2^2,k^2)
  \left(p^{\mu} - \frac{p_2^2-p_1^2}{k^2}\,k^{\mu}\right) 
  \,,\qquad p =p_1 + p_2\,,\quad k = p_2-p_1\,.
\end{align}
The electromagnetic form factor of the spinless target is defined as
the elastic limit of $\Gamma$,
\begin{align}
\label{FFEM}
F_{\gamma}(Q^2) = \Gamma(m^2,m^2,Q^2) =  \C_1(m^2,Q^2,0) 
= Q\int dz\,z^{-2}e^{-\chi(z)}
  K_1(Qz)\,\left|\h{\Phi}_m(z)\right|^2 \,.
\end{align}
Taking into account the tensorial identities
\begin{align}
\V_2^{\mu\nu} + \V_3^{\mu\nu} \pm \V_4^{\mu\nu} \pm \V_5^{\mu\nu}
= \left((p\pm q_2)^{\mu} - q_1^{\mu}\frac{(p\pm q_2)\cdot q_1}{q_1^2}\right)
\left((p\pm q_1)^{\nu} - q_2^{\nu}\frac{(p\pm q_1)\cdot q_2}{q_2^2}\right) \,,
\end{align}
the amplitude \eqref{SCA} can be written in a form which exhibits the
tensorial structure used in \cite{GX09}
\begin{align}
  \label{SCA2}
  \begin{split}
   T^{\mu\nu} &=  e^2\biggl\{2\C_1(m^2,q_1^2,q_2^2)\,\V_1^{\mu\nu} \\
   &\qquad\quad-
   \left(2p_1^{\mu} - q_1^{\mu}\frac{2p_1\cdot q_1}{q_1^2}\right)
   \left(2p_2^{\nu} - q_2^{\nu}\frac{2p_2\cdot q_2}{q_2^2}\right)
   \sum_{n=0}^{\infty}
   \frac{\Gamma(m^2,m_n^2,q_1^2)\,\Gamma^*(m^2,m_n^2,q_2^2)}{s+m^2_n-i\epsilon}
   \\ &\qquad\quad-
   \left(2p_2^{\mu} - q_1^{\mu}\frac{2p_2\cdot q_1}{q_1^2}\right)
   \left(2p_1^{\nu} - q_2^{\nu}\frac{2p_1\cdot q_2}{q_2^2}\right)
   \sum_{n=0}^{\infty}
   \frac{\Gamma(m^2,m_n^2,q_2^2)\,\Gamma^*(m^2,m_n^2,q_1^2)}{u+m^2_n-i\epsilon}
   \biggr\} \,.
\end{split}
\end{align}
As a by-product, the absorptive part of the forward Compton scattering
amplitude reads
\begin{align}
  \label{IM}
  \nonumber
  &\text{Im}\,T^{\mu\nu}(q^2,s) = e^2    
  \left(p^{\mu} + \frac{1}{x}q^{\mu}\right)
  \left(p^{\nu} + \frac{1}{x}q^{\nu}\right)
  \sum_{n=0}^{\infty} \delta(s+m_n^2)\left|\Gamma(m^2,m_n^2,q^2)\right|^2 \,,\\
  &\qquad\qquad\quad\ \approx 
  e^2\left.\left(\px{m_n^2}{n}\right)^{-1}\right|_{m_n^2=-s}
  \left|\Gamma(m^2,-s,q^2)\right|^2
  \left(p^{\mu} + \frac{1}{x}q^{\mu}\right)
  \left(p^{\nu} + \frac{1}{x}q^{\nu}\right) \,,\\
  \nonumber
  &\qquad\qquad
  q_1=q_2=q\,,\qquad p=2p_1=2p_2\,,\qquad x = -\frac{q^2}{p\cdot q} \,.
\end{align}
and do yield, in the hard-wall model, the same structure functions
$F_1=0$ and $F_2$ as found in \cite{PS03}, in the large-$x$ region and
in the one-particle approximation for intermediate states.

More generally, eq.\,\eqref{IM} relates the scaling properties of the
vertex function $\Gamma$ and of the structure function $F_2$ in a
generic soft-wall model. Indeed the function $Qz\,K_1(Qz)$ decreases
monotonically from 1 to 0 and is exponentially small at large
$Qz$. Hence the $z$-dependence of the electromagnetic current can be
roughly approximated as a step function of width $\O(1/Q)$ and
\begin{align}
  \label{overlap}
  \Gamma(m^2,m_n^2,Q^2) \approx \int_0^{1/Q}\frac{dz}{z^3}\,e^{-\chi(z)}
    \,\h{\Phi}(z)\,\h{\Phi}^*_n(z) \,.
\end{align}
One can differentiate three kinematical regimes for the evaluation of
the overlap integral \eqref{overlap}:
\begin{itemize}
\item $Q^2\gg m^2,\ Q^2\gtrsim m_n^2.\quad$ For $z\lesssim 1/Q$ and
  $Q$ large enough at fixed $\Q^2/m_n^2$, $\h{\Phi}(z)$ has the
  asymptotic behavior $\h{\Phi}(z)\sim z^{\Delta}$, where the
  conformal dimension $\Delta$ is the same as in the hard-wall model
  as long as $\chi(z)\rightarrow 0$ when $z\rightarrow 0$. On the
  other hand, since $n$ is a highly excited state, it is legitimate to
  use a WKB approximation for $\h{\Phi}_n(z)$ \cite{PRSW},
  \begin{align}
    \Gamma(m^2,m_n^2,Q^2) \propto C(m_n)
    \left(\frac{1}{Q}\right)^{\Delta}\,F\left(\frac{Q}{m_n}\right) \,.
  \end{align}
  The squared normalization constant $C^2(m_n)$ of $\h{\Phi}_n(z)$ and
  the semiclassical density of states $\frac{\p m_n^2}{\p n}$ have the
  same dependence upon $m_n$.  Therefore the structure function
  $F_2(Q^2,x)$ has the power-law scaling,
  \begin{align}
    \frac{x}{Q^2}F_2(Q^2,x) \propto 
    \left(\frac{1}{Q^2}\right)^{\Delta} F^2(x)
    \,,\qquad x = \frac{Q^2}{Q^2-s} \,.
  \end{align}
\item $Q^2\gg m^2,\ Q^2\gg m_n^2.\quad$ Then both $\h{\Phi}(z)$ and
  $\h{\Phi}_n(z)$ behave as $z^{\Delta}$ for $z\lesssim 1/Q$. It
  follows that the electromagnetic form factor has the power law
  scaling \cite{BRO07},
  \begin{align}
    F_{\gamma}(Q^2) \propto \left(\frac{1}{Q^2}\right)^{\Delta-1} \,.
  \end{align}
  The identity of the asymptotic scaling behavior for the
  electromagnetic form factor and for the structure functions is a
  generic property of the AdS/QCD models that we consider. Such a
  property, which does not even depend upon whatever value of $\Delta$ is
  picked, is certainly difficult to reconcile with a partonic picture.
\item $Q^2\rightarrow 0.\quad$ In that limit the overlap integral
  reduces to the scalar product of $\h{\Phi}$ and $\h{\Phi}_n$. Since
  the asymptotic states $\Phi_{\text{in}}(x,z)$ and
  $\Phi_{\text{out}}(x,z)$ must be eigenstates of the operator
  \eqref{H} with the same mass eigenvalue, all inelastic channels
  decouple when $Q^2=0$ and only the elastic channel
  remains opened. We shall work out some of the consequences in the
  next sections.

  However we can already observe that there is a violation of elastic
  unitarity in the Compton amplitude \eqref{SCA2} which is inherent to
  the $N_c\rightarrow\infty$ approximation involved in the AdS/QCD
  recipes. Indeed the total elastic Compton cross-section is of order
  $e^4$. Hence the imaginary part of the forward amplitude must vanish
  at order $e^2$ in the elastic limit to comply with the optical
  theorem. An absorptive part of the elastic amplitude can be
  generated only by loop effects which are at present beyond the reach
  of the AdS/QCD correspondence.
\end{itemize}

\section{Virtual Compton Scattering}
\label{DVCS}

When the outgoing photon is real, $q_2^2=0$, a mere inspection of
eq.\,\eqref{Vmunu} shows that, in order to cancel the poles in
$q_2^2$, we must have the following relations between the form factors,
\begin{gather}
  \label{DVCSR}
  \begin{split}
  V_1 + (q_1\cdot q_2)V_3 + (p\cdot q_2)V_5 = 0 \,, \\
  (p\cdot q_2)V_2  + (q_1\cdot q_2) V_4 = 0\,.
  \end{split}
\end{gather}
Hence only three Compton form factors remain independent as already
observed in section \ref{COMPTON}. However eqs.\,\eqref{DVCSR} are not
manifestly satisfied by eqs.\,\eqref{SCA} and \eqref{SVCF}. We should
have
\begin{gather}
  \label{DVCSR1}
  2\C_1 + (q_1\cdot q_2)\,\C_+ + (p\cdot q_2)\,\C_- = 0 \,, \\
  \label{DVCSR2}
  (p\cdot q_2)\,\C_+  + (q_1\cdot q_2)\,\C_- = 0\,.  
\end{gather}
The second relation \eqref{DVCSR2} reads explicitly, with the
notations \eqref{notations},
\begin{align}
\begin{split}
  &(p+q_1)\cdot q_2\times \F_1(m^2,q_1^2,0,s) 
  + (p-q_1)\cdot q_2\times \F_1(m^2,0,q_1^2,u) =  \\
  &\iint dz_1\,dz_2\,\,z_1^{-3}e^{-\chi(z_1)}\,\h{\Phi}_i(z_1)\,\times \\
  &\qquad\left(
    -(p_1^2-s) A_1(z_1)\,\h{G}(z_1,z_2,s)\,A_2^{\star}(z_2) +
    (p_2^2-u) A_2^{\star}(z_1)\,\h{G}(z_1,z_2,u)\,A_1(z_2)
  \right) \\
  &\qquad \times\h{\Phi}_f^{\star}(z_2)\,z_2^{-3}e^{-\chi(z_2)}
  = 0 \,.
\end{split}
\end{align}
We now use the completeness relation \eqref{expansion} satisfied by
the Green function $\h{G}$, and take into account the fact that for
the electromagnetic field,
\begin{align}
  \lim_{q_2^2\rightarrow 0} A_2(z) = 1 \,,\qquad
  \lim_{q_2^2\rightarrow 0} \p_{z}A_2(z) = 0 \,.
\end{align}
We integrate each term over $z_2$ and $z_1$ respectively,
\begin{align}
  \begin{split}
  \int dz_2\,z_2^{-3}e^{-\chi(z_2)}\,
  \h{\Phi}_f^{\star}(z_2)\,\h{\Phi}_n(z_2)
  &= C^{\star}(p_2^2,m_n^2) \,, \\
  \int dz_1\,z_1^{-3}e^{-\chi(z_1)}\,
  \h{\Phi}_i(z_1)\,\h{\Phi}_n^{\star}(z_1)
  &= C(p_1^2,m_n^2) \,.
  \end{split}
\end{align}
Hence eq.\,\eqref{DVCSR2} reads
\begin{align}
  \begin{split}
  &-(p_1^2-s)\sum_n\frac{C^{\star}(p_2^2,m_n^2)}{m_n^2+s-i\epsilon}
  \int dz\,z^{-3}e^{-\chi(z)}\,\h{\Phi}_i(z)\,
  \h{\Phi}^{\star}_n(z)\,A_1(z) \\
  &+
  (p_2^2-u)\sum_n\frac{C(p_1^2,m_n^2)}{m_n^2+u-i\epsilon}
  \int dz\,z^{-3}e^{-\chi(z)}\,\h{\Phi}_f^{\star}(z)\,
  \h{\Phi}_n(z)\,A_1(z) = 0 \,.
  \end{split}
\end{align}
Using the orthogonality relations, the identity \eqref{DVCSR2} holds
true exactly only if the virtual Compton scattering is on-shell,
\begin{align}
  p_1^2 = p_2^2 = -m_{n_0}^2\,,\quad\text{for some }n_0\,.
\end{align}
By the same token eq.\,\eqref{DVCSR1} reduces to the definition of
$\C_1$ in \eqref{SVCF},
\begin{align}
  \begin{split}
  &2\C_1(m^2,q_1^2,0) + (p+q_1)\cdot q_2\times \F_1(m^2,q_1^2,0,s) 
  - (p-q_1)\cdot q_2\times \F_1(m^2,0,q_1^2,u) =  \\
  &2\C_1(m^2,q_1^2,0) 
  - 2\int dz\,z^{-3}e^{-\chi(z)}\left|\h{\Phi}_m(z)\right|^2\,A_1(z)
  = 0 \,.
  \end{split}
\end{align}
We can solve for $\C_{\pm}$ in terms of $\C_1$:
\begin{gather}
\begin{split}
  \C_+ = -\frac{2(q_1\cdot q_2)}{(q_1\cdot q_2)^2 - (p\cdot q_2)^2}\,\C_1\,,
  \qquad
  \C_- = \frac{2(p\cdot q_2)}{(q_1\cdot q_2)^2 - (p\cdot q_2)^2}\,\C_1\,, \\
  (q_1\cdot q_2)^2 - (p\cdot q_2)^2 = (q_1-p)\cdot q_2 \times
   (q_1+p)\cdot q_2 = (m^2+s)(m^2+u) \,,
\end{split}
\end{gather}
The VCS amplitude has no absorptive part since $(m^2+s)(m^2+u)$ vanish
only when $q_2=0$ owing to a non-vanishing mass $m$. Therefore the exact
VCS amplitude with $q_2^2=0$ can be written as
\begin{align}
  \label{TDVCS}
  \begin{split}
    T^{VCS}_{\mu\nu} &= e^2\,\C_1(m^2,q_1^2,0)\biggl(2\V_{1,\mu\nu} -
    \frac{2m^2+s+u}{(m^2+s)(m^2+u)}\left(\V_{2,\mu\nu}+\V_{3,\mu\nu}\right) \\
    &\qquad\qquad\qquad\qquad+
    \frac{s-u}{(m^2+s)(m^2+u)}
    \left(\V_{4,\mu\nu}+\V_{5,\mu\nu}\right)\biggr) \,.
\end{split}
\end{align}
The unique Compton form factor reads
\begin{align}
  \label{FF1}
  \C_1(m^2,q_1^2,0) &=
  \int_0^{\infty} dz\,z^{-3}e^{-\chi(z)}
  \left|\h{\Phi}_m(z)\right|^2\,A_1(z) \,,
\end{align}
and is nothing but the electromagnetic form factor of the spinless
target. Note that this formula holds in principle (for on-shell
amplitudes) for any $q_1^2$, spacelike or timelike, if one can make an
analytic continuation in the photon momentum.

The tensorial structure of the amplitude \eqref{TDVCS} is identical to
point-like scalar electrodynamics in the tree level approximation
except for the electromagnetic form factor which encodes all the
internal structure of a spinless particle in this kind of AdS/QCD
models. The threshold theorem imposes that
\begin{align}
\label{NORM}
  \C_1(m^2,0,0) &= \int_0^{\infty} dz\,z^{-3}e^{-\chi(z)}
  \left|\h{\Phi}_m(z)\right|^2 = 1 \,.
\end{align}
The normalization is completely fixed by electromagnetic gauge
invariance and the Hilbert space structure of the classical solutions
in AdS space with appropriate dilaton background.

\section{Bjorken scaling of the DVCS amplitude}
\label{GPD}

It is instructive to understand the consequences of the simplistic form
of the DVCS amplitude \eqref{TDVCS} for a would-be dual picture in
terms of partonic constituents in these kinds of AdS/QCD models.

The Bjorken scaling of the virtual Compton form factors on a
(pseudo)scalar target is usually analyzed in terms of independent
gauge-invariant tensors expressed in the momenta $q=(q_1+q_2)/2$,
$p=p_1+p_2$ and $\Delta=p_2-p_1=q_1-q_2$ with $p_1^2=p_2^2=-M^2$. The
four independent scalar invariant $s$, $u$, $q_1^2$ and $q_2^2$ are
ordinarily traded for $Q^2$, $\Delta^2$, and the scaling variables
$\xi$ and $\eta$ defined by
\begin{align}
\label{BJ}
Q^2 = q^2 \,,\qquad
\xi = -\frac{Q^2}{p\cdot q}\,,\qquad
\eta=-\frac{\Delta\cdot q}{p\cdot q} \,.
\end{align}
The large $Q^2$ expansion of the virtual Compton scattering amplitude
on any target can be described up to twist-three, and with $q_1^2$ and
$q_2^2$ arbitrary, by the tensorial structure \cite{BMKS}
\begin{align}
 \label{TWIST3}
 \begin{split}
 T^{TW3}_{\mu\nu}(q,p,\Delta) &= -\P_{\mu\sigma}g^{\sigma\tau}\P_{\tau\nu}
 \frac{q\cdot W_1}{p\cdot q}
 + \left(\P_{\mu\sigma}p^{\sigma}\P_{\rho\nu}
 + \P_{\mu\rho}p^{\sigma}\P_{\sigma\nu}\right)
 \frac{W^{\rho}_2}{p\cdot q} \\
 &\quad- \P_{\mu\sigma}i\epsilon^{\sigma\tau q\rho}\P_{\tau\nu}
  \frac{A_{1\rho}}{p\cdot q}\,.
 \end{split}
\end{align}
where current conservation is ensured by means of the projector
\begin{align}
  \P_{\mu\nu} = g_{\mu\nu} - \frac{q_{2\mu}q_{1\nu}}{q_1\cdot q_2} \,,
\end{align}
and the transverse component of the momentum transfer is defined by
\begin{align}
  \Delta^{\perp}_{\mu} = \Delta_{\mu} + \eta\,p_{\mu} \,.
\end{align}
The vector $W_{2\rho}$ depends on $W_{1\rho}$ and $A_{1\rho}$ by the
relation ($\epsilon_{0123}=1$),
\begin{align}
W_{2\rho} &= \xi W_{1\rho} - \frac{\xi}{2}\frac{q\cdot W_1}{p\cdot q} p_{\rho}
 + \frac{i}{2}\frac{\epsilon_{\rho\sigma\Delta q}}{p\cdot q} A_{1\sigma} \,.
\end{align}
For a spinless target, the vectors $W_{1\rho}$ and $A_{1\rho}$ are
defined in terms of three generalized form factors
$\H(\xi,\eta,\Delta^2,Q^2)$, $\H_3(\xi,\eta,\Delta^2,Q^2)$ and
$\w{\H}_3(\xi,\eta,\Delta^2,Q^2)$ by the relations
\begin{align}
\begin{split}
 W_1 &= \H\,p + \H_3\,\Delta_{\perp}\,,\quad 
 A_{1\rho} = \frac{i\epsilon_{\rho\Delta p q}}{p\cdot q}\w{H}_3\,.
 \end{split}
\end{align}
When $q_2^2=0$, the VCS relations \eqref{DVCSR} are satisfied and the
generalized form factors $\H$'s are related to the independent form
factors $V_1$, $V_3$ and $V_4$ in \eqref{Vmu} as follows,
\begin{align}
\begin{split}
V_1 &= -\H \,, \\
V_3 &= \frac{1}{\left(2-\frac{\eta}{\xi}\right)}\frac{1}{Q^2}\left(
\left(1-\frac{1}{2-\frac{\eta}{\xi}}\right)\H 
- \xi\w{H}_3\left(2-3\frac{\eta}{\xi}\right)\right) \,, \\
V_4 &= \frac{-\xi}{2-\frac{\eta}{\xi}}\frac{1}{Q^2}
       \left(\H + 2\eta\H_3
  + \xi\w{\H}_3\left(2-\frac{\eta}{\xi}\right)^2\right) \,\\
Q^2 &= \frac{q_1^2}{2}\left(1-\frac{\Delta^2}{2q_1^2}\right) \,.
\end{split}
\end{align}
In perturbative QCD, these form factors can in principle be related,
through factorization, to generalized parton distributions (GPDs).
However, the absence of an absorptive part in the DVCS amplitude
\eqref{TDVCS} is difficult to accommodate with a partonic
interpretation which is based on the convolution of real GPDs with
coefficient functions which contain both a real and an imaginary part.

Specializing to the Bjorken limit of the DVCS amplitude \eqref{TDVCS},
\begin{align} 
\Delta^2 = 0\,,\qquad \xi=\eta=\frac{x_B}{2-x_B}\,,\qquad 
x_B = -\frac{q_1^2}{2p_1\cdot q_1} \,,
\end{align}
one gets for the generalized form factors
\begin{align}
  \begin{split}
  \H &= -2\,\C_1(m^2,2Q^2,0) \,, \\
  \H_3 &= -\w{\H}_3 = \frac{\xi}{1-\xi^2}\,\H 
  = \frac{x_B(2-x_B)}{4(1-x_B)}\,\H\,.
  \end{split}
\end{align}
Therefore the asymptotic behavior in $Q^2$ of the DVCS cross-section
integrated over $t$ and over the azimuthal angle is governed by the
power-law behavior of the electromagnetic form factor. A power-law
behavior in accordance with the dimensional counting rules, e.g. a
scaling dimension $\Delta=2$ for the pion, cannot be consistent with a
partonic interpretation of the DVCS amplitude for spinless hadronic
targets.

\section{Polarizabilities}
\label{POL}

The structure of the VCS amplitude \eqref{TDVCS} and the threshold
theorem, eq.\,\eqref{NORM}, have a still more drastic
consequence. Real Compton scattering on a scalar target in AdS/QCD
models with minimal coupling to the photon is exactly the same as in
point-like scalar electrodynamics in the tree level approximation, a
fact which was observed in the hard-wall model of \cite{GX09}. We
elaborate on the implications for AdS/QCD in this section.

The first consequence is that the static polarizabilities of the scalar
target vanish. Polarizabilities give the corrections to Thompson
scattering which are quadratic in the energy of the photons
\cite{POL}. The amplitude for real Compton scattering off a spinless
particle like the pion,
\begin{align*}
  \gamma(q_1)\,\pi(p_1)\ \longrightarrow\ \gamma(q_2)\,\pi(p_2) \,,
\end{align*}
can be expanded in powers of the energies of the photons near
threshold and reads, in the non-relativistic limit, $\omega_i^2\ll
m_{\pi}^2$, in the laboratory frame and in the Coulomb gauge,
\begin{align}
  A(\gamma\pi\rightarrow\gamma\pi) = 2e^2\v{\epsilon_1}\cdot\v{\epsilon_2}
    + 8\pi m_{\pi}\,\omega_1\omega_2
    \bigl(\alpha_{E}\v{\epsilon_1}\cdot\v{\epsilon_2} + \beta_{M}\,
  (\v{\epsilon_1}\times \h{q}_1)\cdot(\v{\epsilon_2}\times \h{q}_2)\bigr)
  + \cdots \,
\end{align}
where $q_i=\omega_i(1,\h{q}_i)$ and $\v{\epsilon_i}$ are the momentum
and polarization vector of each photon (with
$\h{q}_i^2=\v{\epsilon_i}^2=1$).  $\alpha_{E}$ and $\beta_{M}$ are the
electric and magnetic polarizabilities respectively.  They measure the
linear response of a particle with an internal structure to a small
external electromagnetic perturbation.

The cancellations of the poles in $q_1^2=0$ and $q_2^2=0$ in the
Compton tensor \eqref{Vmunu} impose the following relations between
the Compton form factors,
\begin{align}
V_1 + (q_1\cdot q_2) V_3 = -(p\cdot q_1)V_4 = -(p\cdot q_2)V_5 
= \frac{(p\cdot q_1)(p\cdot q_2)}{q_1\cdot q_2}\,V_2\,.
\end{align}
The most general gauge-invariant real Compton tensor can be written in
terms of the two independent form factors $V_1$ and $V_2$,
\begin{align}
  T^{\mu\nu}_{RCS} &= V_1
  \left(g^{\mu\nu}-\frac{q_1^{\nu}q_2^{\mu}}{q_1\cdot q_2}\right)
  + V_2\left(p^{\mu}-\frac{p\cdot q_1}{q_1\cdot q_2}q_2^{\mu}\right)
  \left(p^{\nu}-\frac{p\cdot q_2}{q_1\cdot q_2}q_1^{\nu}\right) \,.
\end{align}
Since the static polarizabilties $\alpha_E$ and $\beta_M$ are defined
in the laboratory frame, $\v{p_1}= \b{0}$, it is convenient to choose
the Coulomb gauge and impose the conditions $\epsilon_1\cdot p_1 =
\epsilon_2^*\cdot p_1 = 0$. Therefore the contracted real Compton
amplitude can be written as
\begin{align}
  A_{\text{RCS}} = \epsilon_1^{\mu}\,T_{\mu\nu}\,\epsilon_2^{\star\nu} 
  =  V_1\,\v{\epsilon_1}\cdot\v{\epsilon_2}^{\star}
  +\left(V_3-V_2\right)
  (\v{\epsilon_1}\cdot \v{q_2})(\v{\epsilon_2}^{\star}\cdot\v{q_1}) \,.
\end{align}
We can use the identity,
\begin{align}
 (\v{\epsilon_1}\times\v{q_1})\cdot
 (\v{\epsilon_2}\times\v{q_2})
 &= (\v{\epsilon_1}\cdot\v{\epsilon_2})
 (\v{q_1}\cdot\v{q_2}) -
 (\v{\epsilon_1}\cdot\v{q_2})(\v{\epsilon_2}\cdot\v{q_1}) \,,
\end{align}
and relate the electric and magnetic polarizabilities in the lab-frame
to the Compton form factors,
\begin{align}
\label{EM}
\begin{split}
  8\pi m\,\alpha_E &= \left.\pxx{}{\omega_1}{\omega_2}
  \left(V_1+(V_3-V_2)\v{q_1}\cdot\v{q_2}\right)\right|_{\omega_1=\omega_2=0}\,,
  \\ 8\pi m\,\beta_M &= \left.(V_2-V_3)\right|_{\omega_1=\omega_2=0} \,.
\end{split}
\end{align}
In the particular case of \eqref{TDVCS}, in the limit $q_1^2=0$, we
have $V_1 = 2$ and $V_2=V_3$.  Hence $\alpha_E=\beta_M=0$ as expected
for real Compton scattering on a structureless particle.

The point-like nature of the real Compton scattering is a direct
consequence of the minimal coupling between the bulk vector field and
the bulk scalar field in AdS space together with the boundary
condition \eqref{BC}. In order to get non-vanishing polarizabilities
we need to introduce non-minimal couplings between the bulk vector
field and the bulk scalar field in anti-de Sitter space.

It is well-known that the same problematics is encountered in the
calculation of the pion polarizabilities in chiral perturbation theory
($\chi$PT). At lowest-order in the momentum expansion, the pion is
coupled minimally to the electromagnetic field $A_{\mu}$ and the
polarizabilities vanish. The chiral Lagrangian at tree evel can only
predict the $\pi$-$\pi$ scattering lengths. Only the phenomenological
chiral couplings of order 4 can produce non-zero polarizabilities. It
can be shown \cite{DH89} that the predictions at order $p^4$ in the
chiral limit for the electric and magnetic polarizabilities of the
charged pion,
\begin{align}
  \alpha_E = \frac{4\alpha}{m_{\pi}F_{\pi}^2}(L_9^r+L_{10}^r)\,,\qquad
  \alpha_E + \beta_M = 0 \,,
\end{align}
are generated by the four-dimensional effective Lagrangian,
\begin{gather}
  -i\,L_9\,F_{\mu\nu}\tr\left(Q\,D^{\mu}U(D^{\nu}U)^{\dagger}
    + Q\,(D^{\mu}U)^{\dagger}D^{\nu}U\right) 
  + L_{10}\,F_{\mu\nu}F^{\mu\nu}\,
  \tr\left(Q\,U\,Q\,U^{\dagger}\right) \,, \\ \nonumber
  D_{\mu}U = \p{_\mu}U + ie\,A_{\mu}\,[Q,U] \,.
\end{gather}
It is therefore very easy to write down a covariant and
gauge-invariant effective action in the 5D AdS space that can be added
to the minimal action \eqref{AdS} to generate non-vanishing
polarizabilities at the classical level for a charged (pseudo)scalar
particle, e.g.,
\begin{align}
  \label{XPT}
  \begin{split}
  S'_{\text{AdS}}[\Phi,\Phi^*,A^m] &= 
  \int d^4xdz\,\sqrt{-g}e^{-\chi}\biggl(
    ig_1\frac{e}{2}\,F_{mn}(D^m\Phi(D^n\Phi)^* - (D^m\Phi)^*D^n\Phi) \\
    &\qquad\qquad\qquad\qquad\qquad
    + g_2\frac{e^2}{4} F_{mn}F^{mn}\Phi^*\Phi\biggr) \,.
  \end{split}
\end{align}
Of course one could even go one step further in phenomenology and
introduce non-minimal couplings between bulk fields of various spin
and parity, in the spirit of the effective Lagrangian approach.

\section{Conclusion}
\label{END}

We have worked within the bottom-up approach to the AdS/QCD
correspondence and calculated the Compton amplitude with an arbitrary
dilaton background. There is a very recent study \cite{GX09}, within
the approach of \cite{PS03}, which overlaps with ours. There
are however significant differences which make the two papers
complementary. Working in a generic soft-wall model has helped us to
clarify the Lorentz-invariant and gauge-invariant structure of the
Compton amplitude predicted by AdS/QCD. Moreover the structure of the
Compton amplitude does not depend upon the infrared cutoff
parametrized by the dilaton background.

We have found that the minimal coupling of a bulk (pseudo)scalar field
to the electromagnetic current cannot reproduce the expected
low-energy behavior of the Compton amplitude off a spinless composite
charged particle, and produces a too simple structure in the DVCS
kinematical region for a partonic interpretation.

We have pointed out an obvious signature of this failure, namely the
vanishing of the electric and magnetic polarizabilities of the scalar
target. The experimental situation is rather unsatisfactory since the
extraction of the experimental values is model dependent. For instance
the most recent experimental values for the polarizabilities of the
charged pion are,
\begin{align}
  \begin{split}
  (\alpha_E-\beta_M)_{\pi^+} &= 
  (11.6\pm 1.5_{stat}\pm 3.0_{syst}\pm0.5_{mod})\times 10^{-4}\text{fm}^3
  \quad\text{\cite{MAMI}}\,, \\
  (\alpha_E = -\beta_M)_{\pi^+} &= (2.5 \pm 1.7_{stat}\pm 0.6_{syst})
  \times 10^{-4}\text{fm}^3\quad\text{\cite{COMPASS}}\,.
  \end{split}
\end{align}
Even if these values are still imprecise, the inclusion of non-minimal
couplings to the photon is certainly required to obtain a realistic
description of Compton scattering in AdS/QCD at the classical level.
For instance, we note that non-minimal couplings to vector fields
generate five independent Compton form factors, (as opposed to only
three with the minimal coupling we considered in this paper), as
allowed by gauge and Lorentz invariance.

Such couplings appear naturally in chiral perturbation theory.
Besides, an algebra of currents based on chiral symmetry is the
standard framework to describe the hadronic electromagnetic
current. The AdS/QCD models we have examined do not incorporate chiral
flavor symmetry nor vector meson dominance.  There are several
variants of chiral AdS/QCD models \cite{SS04,EKSS05,RP05,HS05} and it
is not the purpose of the present work to commit to one of
them. Nethertheless we have identified a bare-bones effective anti-de
Sitter action that can contribute to the polarizabilities in many
chiral models.

In any case it should now be clear that the calculation, and the
precise measurement, of the hadronic polarizabilities is a selective
testing ground for the AdS/QCD correspondence.

\subsection*{Acknowledgements}

We wish especially to thank B.~Pire for inspiring discussions and very
useful comments about the manuscript. We also wish to thank
V.~Bernard, J.P.~Lansberg, B.~Moussallam, T.N.~Pham, U.~Reinosa,
L.~Szymanowski and B.~Xiao for very interesting discussions. C.M.  is
supported by the European Commission under the FP6 program, contract
No.  MOIF-CT-2006-039860. This work is partly supported by the
ANR-06-JCJC-0084.

\subsection*{Appendix:\quad Explicit formulas in the hard-wall model}

The hard-wall model is defined by the absence of dilaton background,
$\chi(z)=0$, and by imposing a Dirichlet boundary condition on the
massive fields at some finite cutoff in $z$. Then the plane-wave
solution for a massive scalar field reads (in the timelike region),
\begin{align}
  \Phi(x,z) &= C_{\Delta-1}(m)\,e^{ip\cdot x}z^2\,
  J_{\Delta-2}\left(m\,z\right)
  \,,\quad \Delta=2\pm\sqrt{m_S^2+4} \geq 1\,,\quad p^2=-m^2<0 \,, 
\end{align}
(this is not the most general admissible solution for $1\leq\Delta\leq
3$).  The normalization constants,
\begin{align}
  \label{normalization}
  C_{\Delta-1}(m) = \sqrt{2}\,\Lambda\,
  J_{\Delta-1}^{-1}\left(\frac{m}{\Lambda}\right)\,,
\end{align}
are defined by requiring,
\begin{align}
  \int_0^{1/\Lambda}dz\,z^{-3}|\Phi(x,z)|^2 &= 1\,,
  \quad J_{\Delta-2}\left(\frac{m}{\Lambda}\right) = 0 \,.
\end{align}
The scalar propagator reads,
\begin{gather}
 \h{G}(z_1,z_2,-m^2) = \sum_n C^2_{\Delta-1}(m_n)
 \frac{z_1^2J_{\Delta-2}(m_nz_1)\,z_2^2J_{\Delta-2}(m_nz_2)}
 {m^2-m_n^2+i\epsilon}\,,\\
 m_n = \zeta_{\Delta-2,n}\Lambda \,,
\end{gather}
where $\zeta_{\nu,n}$ are the zeroes of the Bessel function
$J_{\nu}(z)$.  When $\Lambda\rightarrow 0$, the scalar propagator
becomes,
\begin{align}
  \h{G}(z_1,z_2,-m^2) \underset{\Lambda\rightarrow 0}{\approx}
  (z_1z_2)^2\int_0^{\infty}d\mu\,\mu\,
  \frac{J_{\Delta-2}(\mu z_1)\,J_{\Delta-2}(\mu z_2)}
  {m^2-\mu^2+i\epsilon} + \O(\Lambda) \,.
\end{align}
\vskip 0.2cm 
Plugging the explicit wave-functions into \eqref{FF1} one
gets for the scalar DVCS form factor,
\begin{align}
  \C_1(m^2,Q^2,0) = C^2_{\Delta-1}(m)Q\int_0^{1/\Lambda} dz\, z^2\,
  J^2_{\Delta-2}(mz)\,K_1(Qz) \,.
\end{align}
As long as $\Delta>1$ and $Q\gg\Lambda$, we can set $\Lambda=0$ in the
integration domain and use the integral formula,
\begin{align}
  \C_1(m^2,Q^2,0) &\simeq 2(\Delta-1)\frac{C^2_{\Delta-1}(m)}{m^2}\times
  \left(\frac{m^2}{Q^2}\right)^{\Delta-1}\times
  \frac{(1-w)^{2\Delta}}{(1-w^2)^2}
 \left(1 + \frac{1}{\Delta-1}\frac{2w^2}{1-w^2}\right) \,,
\end{align}
where $w$ is defined by
\begin{align}
  w =  1+\frac{Q^2}{2m^2} -
  \sqrt{\left(1+\frac{Q^2}{2m^2}\right)^2-1}\,.
\end{align}
Noting that
\begin{gather}
 w \underset{Q^2\rightarrow\infty}{\approx}
 \frac{m^2}{Q^2}\left(1+\O\left(\frac{m^2}{Q^2}\right)\right)
 \longrightarrow 0\,,
\end{gather}
we obtain the leading large $Q^2$ behavior of the DVCS form factor
found in \cite{GX09},
\begin{align}
  \C_1(m^2,Q^2,0) &= 2(\Delta-1)\frac{C^2_{\Delta-1}(m)}{m^2}\times
  \left(\frac{m^2}{Q^2}\right)^{\Delta-1} \,.
\end{align}

\end{document}